\title{\LARGE \bf Multidimensional Hermite polynomials \\
and photon distribution  \\ for polymode mixed light\\[7mm] }

\author{ V.~V.~Dodonov, O.~V.~Man'ko\\
{\it Lebedev Physics Institute, Moscow, Russia}\\[3mm]
and V.~I.~Man'ko\\{\it Dipartimento di Scienze Fisiche,}\\
{\it Universita di Napoli "Federico II"}\\
{\it and I.N.F.N., Sez. di Napoli} }

\begin{document}
\maketitle

\vfill

INFN-NA-IV-93/36 \hfill DSF-T-93/36

\newpage

\begin{abstract}
For N-mode light described by the Wigner function of generic Gaussian
form the photon distribution function is obtained explicitly and expressed
in terms of Hermite polynomial of $2N$ variables
with equal pairs of indices.The mean values and dispersions
of photon numbers are obtained for this generic mixed state.Generating
function for photon distribution is obtained explicitly.
The expression for $N$-mode photon distribution function for
squeezed photon number states in terms of Hermite polynomials
of $2N$ variables and for squeezed coherent states in terms of
Hermite polynomials of $N$ variables is discussed.

\end{abstract}

\newpage

\section{Introduction}

The problem of finding the photon distribution function for the one-mode
electromagnetic field state described by generic Gaussian Wigner function
has been considered in our previous work [1] where this function was
expressed in terms of Hermite polynomials of two variables with equal indices.

The aim of the work is to show that the photon distribution function
of the N-mode field state described by the generic Gaussian Wigner
function may be calculated explicitly and to express it in terms of
Hermite polynomials of $2N$ variables with equal paires of indices.We will also
discuss the expression of photon distribution function for the $N$-mode light
in squeezed number state which may be given in terms of Hermite polynomials
of $2N$ variables[2].All these expressions are the same as for the
probabilities
distribution functions for parametric polymode oscillator
excited by an external force obtained in [3],[4].

We will clarify the physical meaning of the parameters on which depends the
photon distribution function using the Wigner representation for the
field density operator since in this representation it is obvious that
there are $2N^{2}+3N$ parameters in generic case of Gaussian Wigner function.

Our goal is to express
the photon distribution function in terms of these parameters explicitly.

\section{Gaussian Wigner function}

The mixed squeezed state of the $N$-mode
light with density operator $\hat \rho $ is
described by Wigner function $W({\bf p},{\bf q})$ of the generic Gaussian form
which contains $2N^2+3N$ real parameters.$2N$ parameters are mean
values of light quadratures $<{\bf p}>$ and quadratures
$<{\bf q}>$  and other $2N^2+N$ parameters are
matrix elements of the real symmetric dispersion
matrix $m$ with four $N$-dimensional block matrices three of which are
\begin {eqnarray}
m_{11}=\sigma _{\bf p},\\
m_{12}=\sigma _{\bf p\bf q},\\
m_{22}=\sigma _{\bf q}.
\end {eqnarray}
Below we will use the invariant parameters
\beq
T=Tr~{m}
\eeq
and
\beq
d=\det m.
\eeq
Also we will use the other invariant coefficients of the polynomial
\beq
P(x)=\det (m-x\bf 1)
\eeq
where $\bf 1$ is $2N$-dimensional identity matrix.

We will introduce the notations
\beq
{\bf Q}=({\bf {p}},{\bf {q}})
\eeq
where $2N$-dimensional vector $\bf Q$ consists of $N$ components of light
quadratures
$p_{1},...,p_{N}$ and $N$ components of quadratures $q_{1},...,q_{N}$.
The generic gaussian Wigner function has the form (see,for example,[4])
\beq
W({\bf {p}},{\bf {q}})
=d^{-\frac{1}{2}}\exp [-(2)^{-1}[({\bf {Q-<Q>}})m^{-1}({\bf {Q-<Q>}})].
\eeq
The parameters $<\bf p>$ and $<\bf q>$ are given by the formulae
\begin {eqnarray}
<{\bf p}>=Tr~ \hat \rho \hat {\bf p} ,\\
<{\bf q}>=Tr~ \hat \rho {\hat {\bf q}},
\end {eqnarray}
where the operators $\hat {\bf p}$ and $\hat {\bf q}$
are the quadrature components of
photon creation $\bf a\dag $ and the annihilation ${\bf a}$ operators
\begin {eqnarray}
\hat {\bf p}=\frac{{\bf a}-{\bf a\dag}}{i\sqrt 2},\\
\hat {\bf q}=\frac{{\bf a}+{\bf a\dag}}{\sqrt 2}.
\end {eqnarray}
Due to the physical meaning of the dispersions the diagonnal elements
of the matrix $m$  must be nonnegative numbers,so the invariant parameter
$T$ Eq.(4) is a positive number.Also the determinant $d$ Eq.(5)
of the dispersion
matrix must be positive.In fact the matrix $m$ must be
positive-definite.The function (8) if one does not precise that it
is Wigner function of a quantum system may be also concidered formally as
classical Gaussian distribution function.Then the above constrains of the
parameters domain take place for such formal distribution function.And
in classical version of the function (8) there is no others constrains.But
for quantum Wigner function the uncertainty relation provides additional
restriction and the domain of permitted values of the matrix elements of
the dispersion matrix is smaller for quantum Wigner function than for
classical Gaussian distribution function in phase space of a particle.

\section{Photon distribution function}

To obtain the photon distribution function we have to calculate the probability
$P _{\bf n}$ to have $\bf n$
photons in the state with the density operator $\hat \rho$.
Here the vector $\bf n$ has $N$ components $n_i$ which are nonnegative
integers.
This probability is given by the formula
\beq
P_ {\bf n}=Tr~ {\hat \rho}|{\bf n}><{\bf n}|,\\n_{i}=0,1,2,...~~~i=1,2,...,N
\eeq
where the number states $|{\bf n}>$ are the eigenstates of the number operator
$\bf a\dag a$
\beq
{\bf a\dag a|n>=n|n>}.
\eeq
The function $P_ {\bf n}$ may
be obtained if one calculates the generating function
for the matrix elements $\rho _{\bf mn}$ of the density operator $\hat \rho$ in
the Fock basis.This generating function is the matrix element of the density
operator in the coherent state basis
\beq
< {\bf\beta} | \hat \rho | {\bf\alpha} > =
\exp ( \frac {-|{\bf\alpha}|^2}{2} - \frac {|{\bf\beta}|^2}{2} )
 \sum_{{\bf m,n=0}}^{\infty }
\frac{ ({\bf \beta }^{*})^{{\bf m}} {\bf \alpha}^{\bf n}}
{   ( {\bf m!n!} )^{\frac {1}{2} }   }
\rho_{{\bf mn}}
\eeq
Here and below we use notations:$\alpha $and $\beta $ are $N$-dimensional
vectors with complex components and
\begin {eqnarray}
{\bf n}!=n_{1}!n_{2}!...n_{N}!,\\
{\bf \alpha }^{\bf n}=\alpha _{1}^{n_1}\alpha _{2}^{n_2}...\alpha _{N}^{n_N}
\end {eqnarray}
and
\beq
\sum_{{\bf m,n=0}}^{\infty}=\sum_{m_{1}=0}^{\infty}...\sum_{m_{N}=0}^{\infty}
\sum_{n_{1}=0}^{\infty}...\sum_{n_{N}=0}^{\infty}.
\eeq
We have
\beq
P_ {\bf n}=\rho _{{\bf nn}}.
\eeq
The function $<{\bf \alpha}|\hat \rho |{\bf \alpha}>$ is the Q-function of
the system with the density operator $\hat \rho $.
The $N$-mode coherent
state $|{\bf \alpha}>$ is the normalized eigenstate of the
annihilation operator
\beq
{\bf a}|{\bf \alpha}>={\bf \alpha}|{\bf \alpha}>.
\eeq
In terms of Wigner function the density operator in the coherent state
representation has the form of $2N$-dimensional overlap integral [4]
\beq
<{\bf \beta}|\hat \rho|{\bf\alpha}>=
\frac {1}{({2\pi })^{N}}\int W({\bf p},{\bf q})W_{{\bf\alpha}{\bf \beta} }
({\bf p},{\bf q})d{\bf p}d{\bf q},
\eeq
where the function $W_{{\bf \alpha} {\bf\beta }}({\bf p},{\bf q})$
is the Wigner function of the
operator $|{\bf\alpha}><{\bf\beta}|$.It has the form [4]
\beq
W_{{\bf\alpha} {\bf\beta }}({\bf p},{\bf q})=2^{N}
\exp [-\frac {|{\bf\alpha} |^2}{2}-\frac {|{\bf\beta}|^2}{2}-
{\bf\alpha} {{\bf\beta}}^{*}-{\bf p}^{2}-{\bf q}^{2}+
\sqrt 2{\bf\alpha} ({\bf q}-i{\bf p})+\sqrt 2{{\bf\beta}}^{*}
({\bf q}+i{\bf p})].
\eeq
Let us introduce the $2N$-dimensional vector
\beq
{\bf \gamma}=({\bf \beta }^{*},{\bf \alpha })
\eeq
which is composed from two $N$-dimensional vectors.Let us introduce also
$2N$- dimensional unitary matrix $u$ which has four $N$-dimensional block
matrices
\beq
u_{11}=-u_{12}=-iu_{21}=-iu_{22}=-\frac {i}{\sqrt 2}{\bf 1}
\eeq
and ${\bf 1}$ is $N$-dimensional identity matrix.

Since the integral given by Eq.(21) is the Gaussian one it may be easily
calculated.
So,we have
\beq
<{\bf \beta}|\hat \rho|{\bf\alpha}>=P_{0}
\exp (-\frac{|{\bf \gamma }|^2}{2} )
\exp [-\frac{1}{2}{\bf \gamma }R{\bf \gamma }+{\bf \gamma }R{\bf y}]
\eeq
where the symmetric $2N$-dimensinal matrix $R$
has the matrix elements expressed in terms
of the dispersion matrix $m$ as follows
\beq
R=-2u^{-1}(\frac {m^{-1}}{2}+1)^{-1}u^{*}+\Sigma _{x}
\eeq
The matrix $ \Sigma _{x}$ is $2N$-dimensional analog of Pauli matrix.
The $2N$-dimensional vector ${\bf y}$ is given by the relation
\beq
{\bf y}=-2u^{t}(1-2m)^{-1}<{\bf Q}>
\eeq
where the matrix $u^t$ is transposed one and the factor $P_{0}$ has the form
\beq
P_{0}=[det (m+\frac {1}{2})]^{-\frac {1}{2}}
\exp[-<{\bf Q}>(2m+1)^{-1}<{\bf Q}>]
\eeq
In the case of quadrature means equal to zero the probabability to have
no photons depends on $2N-1$ parameters which is the number of
invariant coefficients
of the polynomial (6).

The function $\exp[\frac {|{\bf \gamma }|^{2}}{2}]
<{\bf \beta}|\hat \rho|{\bf\alpha}>$ is the generating function for the
matrix elements of the density matrix in the photon number state basis.
This function coincides with the generating function for Hermite polynomials
of $2N$ variables[4].
The probability to have no photon is equal to $P_0$ given by formula (28)
Comparing the formula (25) with the generating function for Hermite
polynomials of $2N$ variables [5]
we obtain for the photon distribution function $P_{{\bf }n}$ the expression
\beq
P_{{\bf n}}=P_{0}\frac{H_{{\bf nn}}^{\{R\}}({\bf y})}{{\bf n}!}.
\eeq
Here the matrix $R$ determining the Hermite polynomial is given by the
formulae (26) and  argument of Hermite polynomial is given by the
expression (27).The expression (29) is the partial case of the matrix
elements of the density operator in Fock states basis obtained in [2]
by canonical transform method.

\section{Average values for multidimensional Hermite polynomials}

In [1] for one-mode light
was obtained the expression for mean value of photon number
in the mixed gaussian state and also the expression for dispersion
of photon number was found.In this section we will calculate these
quantities for polymode mixed gaussian light.First we will derive some
formulae for multidimensional Hermite polynomials
We start from the generating function for ``diagonal''
multidimensional Hermite polynomials [1] which gives in our case the
generating function for photon distribution (29)
\begin{eqnarray}
\sum_{n_1=0}^{\infty}\ldots\sum_{n_N=0}^{\infty}\frac {
\lambda_1^{n_1}}{n_1!}\frac {\lambda_2^{n_2}}{n_2!}\ldots\frac {\lambda_
N^{n_N}}{n_N!}H_{n_1n_2\ldots n_Nn_1n_2\ldots n_N}^{\{R\}}(R^{-1}
{\bf z})= \nonumber \\
=i^N[\det(\Lambda R+\Sigma_x)]^{-\frac 12}\exp[\frac 12{\bf z}
(R+\Lambda^{-1}\Sigma_x)^{-1}{\bf z}],
\end{eqnarray}
where ${\bf z}=(z_1,z_2,...z_{2N})$, the $2N$x$2N$ matrix $\Sigma_
x$ is the
$2N$-dimensional analog of the Pauli matrix $\sigma_x$.
The  diagonal $2N$x$2N$-
matrix $\Lambda$ has the form

\[\Lambda =\sum_{j=1}^N\lambda_j{\cal M}_j,\]
where each matrix ${\cal M}_j$ has only two nonzero elements:
\[({\cal M}_j)_{jj}=({\cal M}_j)_{j+N,j+N}=1,\]
while $\lambda_j$ may be arbitrary complex numbers.  Let us define the
"average values" $<n_1^{\alpha_1}n_2^{\alpha_2}\cdots n_N^{\alpha_N}>$
by the relation

\begin{eqnarray}
\sum_{n_1=0}^{\infty}\ldots\sum_{n_N=0}^{\infty}n_1^{\alpha_1}
n_2^{\alpha_2}\cdots n_N^{\alpha_N}\frac {H_{n_1n_2\ldots n_Nn_1n_
2\ldots n_N}^{\{R\}}({\bf y})}{n_1!n_2!\cdots n_{_{}N}!}= \nonumber \\
=\mbox{$<
n_1^{\alpha_1}n_2^{\alpha_2}\cdots n_N^{\alpha_N}>$}\sum_{n_1=0}^{
\infty}\ldots\sum_{n_N=0}^{\infty}\frac {H_{n_1n_2\ldots n_Nn_1n_
2\ldots n_N}^{\{R\}}({\bf y})}{n_1!n_2!\cdots n_{_{}N}!}.
\end{eqnarray}

Evidently, they can be obtained by differentiating the generating
function over $\lambda_j$ and putting in the
final expression $\lambda_k=1$ for all
$k=1,2,\ldots N$.  It is not difficult to differentiate the matrix defining
the quadratic form in the exponential function, if one takes into
account the relation
$$\frac{\partial\Lambda}{\partial\lambda_j}={\cal M}_j.$$
To differentiate the matrix determinant standing in the
preexponential factor we use the known formula
$$\frac{d}{dt} \log\det A(t)={\rm T}{\rm r}(A^{-1} \frac{dA}{dt}). $$
The first derivatives of the generating function yield the first
order average values

\beq
<n_j>=
\frac{1}{2} {\rm T} {\rm r}
\left[ {\cal M}_j {\cal S} \right]
-1 + \frac{1}{2} {\bf y}R
\mbox{$ {\cal S}{\cal M}_j{\cal S}\Sigma_x R {\bf y} $},
\eeq

where matrix ${\cal S}$ is given by
$${\cal S}=(\Sigma_xR+{\bf 1})^{-1}.$$

Note that matrix ${\cal M}_j$ commutes with matrix $\Sigma_x$, but does not
commute, in general, with matrix R.  However, in the one-mode
case, when all matrices have only two rows and columns, ${\cal M}_
j={\bf 1}$,
and the formulas are simplified.

The covariances
\[\sigma_{jk}=<n_jn_k>-<n_j><n_k>\]
can be found in the same manner: one should calculate  the second
derivatives of the generating function and subtract the products of
the first order averages. For $j\neq k$, taking into account the relations

\[{\cal M}_j{\cal M}_k={\bf 0},\qquad {\cal M}_j\Sigma_x=\Sigma_x
{\cal M}_j,\]
we get
\beq
\sigma_{jk}(R,{\bf y})=
\frac 12{\rm T}{\rm r}\left[{\cal M}_j{\cal S}
{\cal M}_k{\cal S}\right]+
\frac 12 {\bf y}R{\cal S}{\cal M}_j{\cal S}
{\cal M}_k{\cal S}\Sigma_xR{\bf y}.
\eeq
The second derivative of the generating function with respect to
the same argument $\lambda_j$ gives $<n_j(n_j-1)>$. Taking into account that
${\cal M}_j^2={\cal M}_j$, we obtain the following expression for the variance:
\beq
\sigma_{jj}(R,{\bf y})=\frac 12{\rm T}{\rm r}\left[{\cal M}_j{\cal S}
({\cal M}_j{\cal S}-1)\right]+\frac 12{\bf y}R(2{\cal S}{\cal M}_
j-1){\cal S}{\cal M}_j{\cal S}\Sigma_xR{\bf y}.
\eeq
Since the photon distribution function (29) is expressed in terms of
multidimensional Hermite polynomials it is easy to see that the photon numbers
means and the dispersions are given by formulae obtained above but with
argument
$R$ and ${\bf y}$ given by relations (26),(27).

\section{Sqeezed coherent and sqeezed number states}

In this section we will discuss the photon distributions for polymode light
in squeezed coherent states and squeezed number states.The mathematical
structure of the formulae for these distributions is identical to
transition probabilities formulae for polymode parametric oscillator
expressed in [3] in terms of Hermite polynomials of several variables.
We will give these expressions following [4].Let us inntroduce the
linear canonical transform of photon creation and annihilation operators
of the form
\beq
{\bf B}=S{\bf A}+{\bf D},~~~~~~~~~{\bf A}=({\bf a},{\bf a}\dag)
\eeq
where the $2N$x$2N$-matrix $S$ belongs to symplectic group and has the
four $N$x$N$-matrices
\beq
S_{11}=\zeta =S_{22}^{*},~~~~S_{12}=\eta =S_{21}^{*}.
\eeq
The complex $2N$-vector ${\bf D}=({\bf d},{\bf d}^{*})$ produces the shift
of photon quadrature components.Here ${\bf d}$ is complex $N$-vector.
The photon distribution function may be expressed in case of squeezed
polymode light in terms of these parameters.

For all the pure states which are described by the gaussian Wigner function
the matrix $R$ depends on smaller number of
parameters than in case of generic mixed gaussian state.
The parameter which is used to distinguish
the mixed states is the following
\beq
\mu=Tr~\hat \rho ^{2}.
\eeq
For $N$-dimensional system with gaussian Wigner function the parameter
is determined by the determinant of the dispersion matrix $m$.It is
\beq
\mu=\frac{1}{(2\pi )^{N}}\int W^{2}({\bf p},{\bf q})d{\bf p}d{\bf q}
=\frac{1}{2^{N}\sqrt d}.
\eeq
For $d=\frac{1}{4^{N}}$ the parameter $\mu =1$ that corresponds to the pure
states.The obvious relation
\beq
Tr~\hat \rho ^{2}\leq Tr~\hat \rho =1,~~~~d\geq \frac{1}{4^{N}}
\eeq
gives the uncertainty relation which in one-dimensional case
is the Schrodinger uncertainty relation [6]. The photon distribution function
for pure gaussian state may be expressed in terms of Hermite polynomials of
$N$ variables.It is related to the minimisation of analog of Schrodinger
relation for polymode case.

So for squeezed coherent
light in state $|\beta ,s>$ which is the eigenstate of operator ${\bf B}$
 (35) the photon distribution function is
\beq
\frac {P_{{\bf n}}}{P_{0}}=\frac {|H_{{\bf n}}^{\{R\}}({\bf y})|^2}{{\bf n}!}.
\eeq
where the $N$x$N$-matrix $R$ is
expressed in terms of symplectic matrix $S$ (35),(36)
\beq
R=\zeta ^{-1}\eta
\eeq
and argument ${\bf y}$ of
the Hermite polynomial of $N$ variables is determined by
the  inhomogeneous
symplectic transform and the label of squeezed coherent state
\beq
{\bf y}=\eta ^{-1}(\beta -{\bf d}).
\eeq
The term in denominator $P_0$ is the probability to have no photon.

Thus we have shown that photon distribution function of pure gaussian
state may be described either by the formula (29) or by the formula (40).
It means that there must exist an interesting new relation connecting Hermite
polynomials of $2N$ variables with the Hermite polynomials of $N$ of
variables.

The dispersion matrix $m$ in pure gaussian state $|\beta,s>$
may be expressed in terms
of symplectic transform $S$

\beq
m=-\frac {1}{2}u\Sigma _{y}S^{t}\Sigma _{x}S\Sigma _{y}u^{t}
\eeq
where the $2N$x$2N$-matrix $\Sigma _y$ is multidimensional analog
of Pauli matrix $\sigma_y$.So if the matrix $2N$x$2N$-matrix $R$ in
formula (26) is calculated
in terms of above matrix $m$ (43) in which the components of matrix $S$
(36) are taken and if we use same components
for the $N$-matrix $R$ (41) we have equality
\beq
{H_{{\bf nn}}^{\{R\}}({\bf Y})}=|H_{{\bf n}}^{\{r\}}({\bf y})|^2
\eeq
Here in right hand side of the equality we replaced the notation
to be small $r$ but this matrix $r=\zeta ^{-1}\eta $ as given by (41).
We also replaced argument of Hermite polynomial in left hand side of
the equality by ${\bf Y}$ and this argument is given by the formula (27)
in which
\beq
<{\bf Q}>=uS^{-1}(<{\bf B}>-{\bf D})
\eeq
and $<{\bf B}>=(\beta,\beta ^{*})$.The argument ${\bf y}$ is given by (42).

If we have the squeezed number state $|{\bf m}>$ the photon distribution
function is given by the formula
\beq
\frac {P_{{\bf n}}}{P_{0}}=\frac {|H_{{\bf mn}}^{\{R\}}({\bf L})|^2}{{\bf n}!
{\bf m}!}.
\eeq
Here the symmetric
$2N$x$2N$-matrix $R$ is expressed in terms of symplectic matrix $S$ and three
its $N$x$N$-block matrices are

\beq
R_{11}=\zeta ^{-1}\eta,~~~R_{12}=-\zeta ^{-1},~~~R_{22}=-\eta ^{*}\zeta ^{-1}.
\eeq

The $2N$-vector ${\bf L}$ which is the argument of Hermite polynomial of
$2N$ variables is given by the formula
\beq
{\bf L}=R^{*}{\bf X},~~~~~~~~{\bf X}=({\bf x}_{1},{\bf x}_{2})
\eeq
and $N$-components vectors in this formula are
\beq
{\bf x}_1=-\zeta ^{-1}{\bf d},~~~~{\bf x}_2={\bf d}^{*}-
\eta ^{*}\zeta ^{-1}{\bf d}.
\eeq
The behaviour of the distributions may be wavy function of photon
numbers as it is in one-mode case[7],[8].

Using the sum rules (32) we can express mean values of photons in each
mode in the form
\beq
<n_{j}>=\frac {1}{2}(\sigma _{p_{j}p_{j}}+\sigma _{q_{j}q_{j}}-1)+
\frac {1}{2}(<p_{j}>^{2}+<q_{j}^{2}).
\eeq
Using the sum rules (34) we obtain for photon number variances the
expression
\beq
\sigma _{n_{j}}=\frac {1}{2}(T_{j}^{2}-2d_{j}-\frac {1}{2})+
<Q_{j}>m_{j}<Q_{j}>
\eeq
where $T_j$ and $d_j$ are the trace and the determinant of
the photon quadrature $2$x$2$-dispersion matrix  $m_j$ of the j-th mode only
and the 2-vector $Q_{j}$ has components $p_{j},q_{j}$.
The correlations of photon numbers in different modes may be expressed
analogously.

\section{Acknoledgements}
One of us (V.I.M.)thanks INFN and University of Napoli "Federico II"
for the hospitality.

\begin {center}
{\LARGE\bf References}\\
\end {center}

[1] V.V.Dodonov,O.V.Mam'ko and V.I.Man'ko,Photon distribution for
one-mode mixed light with generic gaussian Wigner function,University of Napoli
preprint
INFN-NA-IV-93/35,DSF-T-93/35(1993).

[2] V.V.Dodonov,V.I.Man'ko and V.V.Semjonov,Nuovo Cimento B {\bf 83},145(1984),
see also V.I.Man'ko in Proceedings of Workshop "Harmonic Oscillators"
,University of Maryland,March 25-28,1992

[3] I.A.Malkin,V,I.Man'ko and D.A.Trifonov,J.Math.Phys,{\bf 14},576(1973).

[4] V.V.Dodonov and V.I.Man'ko "Invariants and evolution of
nonstationary quantum systems",Proceedings of Lebedev Physical institute
{\bf 183},ed.by M.A.Markov,Nova Science Publishers,
Commack,N.Y.(1989).

[5] Bateman Manuscript Project:Higher Transcendental Functions,
edited by A.Erdely (McGraw-Hill,New York,N.Y.(1953).

[6] E.Schrodinger,Ber.Kgl.Akad.Wiss.Berlin,296(1930).

[7] W.Schleich and J.A.Wheeler,J.Opt.Soc.Am.B {\bf 4},1715(1987).

[8]  V.V.Dodonov,A.B.Klimov and V.I.Man'ko,Phys.Lett.A {\bf 89},555(1990).

\end{document}